\documentclass[%
 aip,
 amsmath,amssymb,
preprint,%
]{revtex4-1}

\usepackage{graphicx}% Include figure files
\usepackage{dcolumn}% Align table columns on decimal point
\usepackage{bm}% bold math
\usepackage[utf8]{inputenc}
\usepackage[T1]{fontenc}
\usepackage{mathptmx}
\usepackage{makecell}
\usepackage{xcolor}

\usepackage{amsmath}
\usepackage{amsfonts}
\usepackage{amssymb}
\usepackage{natbib}

\newcommand{\velamp}{\boldsymbol{\mathsf{U}}}
\newcommand{\velampx}{{\mathsf{U}_1}}

\newcommand{\pressamp}{\mathsf{P}}

\begin{document}

% \preprint{AIP/123-QED}

\title[]{Quasi-periodically developed flow in channels with arrays of in-line square cylinders}

\author{G. Buckinx}%
\email{geert.buckinx@vito.be.}
\affiliation{%
VITO, Boeretang 200, 2400 Mol, Belgium%\\This line break forced% with \\
}%
\affiliation{%
EnergyVille, Thor Park, 3600 Genk, Belgium%\\This line break forced% with \\
}%
\affiliation{ 
Department of Mechanical Engineering, KU Leuven, Celestijnenlaan 300A, 3001 Leuven, Belgium%\\This line break forced with \textbackslash\textbackslash
}%
\author{A. Vangeffelen}
\affiliation{ 
Department of Mechanical Engineering, KU Leuven, Celestijnenlaan 300A, 3001 Leuven, Belgium%\\This line break forced with \textbackslash\textbackslash
}%
\affiliation{%
EnergyVille, Thor Park, 3600 Genk, Belgium%\\This line break forced% with \\
}%

\date{\today}% It is always \today, today,
             %  but any date may be explicitly specified

\begin{abstract}
In this stub article, we show that laminar quasi-periodically developed flow is characterized by velocity and pressure modes which decay exponentially along the main flow direction.
As the amplitudes of these modes exhibit streamwise periodicity, they can be determined on a single transversal row of the array.
Their shape and corresponding decay rate are governed by an eigenvalue problem which generalizes the Orr-Sommerfeld equation for quasi-developed Poiseuille flow.
By means of full-scale channel flow simulations, we numerically investigate the onset point, extent, eigenvalues and perturbation sizes of quasi-periodically developed flow in channels with equidistant in-line square cylinders, assuming a parabolic inlet velocity profile.
In particular, the dependence on the mass flow rate, the aspect ratio and height of the channel is discussed,  considering three porosities of the cylinder array (0.75, 0.89 and 0.94), and Reynolds numbers from 20 to 300, based on a reference length of twice the channel height.
It is observed that the region of quasi-periodically developed flow covers a large part of the region of flow development in the channel.
Therefore, the corresponding eigenvalues explain largely the observed scaling laws for the onset point of periodically developed flow.

\textbf{Key words}: Developing Flow, Entrance Flow, Eigenvalue Problem, Micro-and Mini-Channels, Periodically Developed flow, Stationary Perturbations
\end{abstract}

\maketitle

\section{\label{sec:Introduction}Introduction}
% Following/Since the introduction/In line with
Motivated by the development of compact heat transfer devices with periodic fin arrays and other periodic heat transfer surfaces \citep{KaysLondon1984, ShahLondon1978}, researchers have investigated different aspects of flow in channels with periodic solid structures.
To give a few examples, we mention here the characterization of the flow regimes and pressure drop in arrays of pin fins, wavy fins, and tube bundles, explored by e.g. \citet{Lawson2011, Junqi2007, Launder1978}.
Due to the recent advances in microfluidic devices \citep{KosarMishraPeles2005} and ordered microporous materials \citep{Zargartalebi2020}, interest in the topic has revived, especially for laminar flow conditions.
The reason is that these applications often employ hundreds of circular or square cylinders in a periodic configuration, with a diameter of 10 to 1000 $\mu$m \citep{SiuHoQuPfefferkorn2007, MohammadiKosar2018, KosarPeles2007}.
As a result of those small dimensions, the flow inside the former  cylinder arrays is characterized by low to moderate Reynolds numbers, and tends to remain laminar and steady.
This observation also applies to the many other types of solid structure arrays encountered in microfluidic devices, like for instance offset strip fins \citep{Vangeffelen2021, Vangeffelen2022}.

Already since the experiments of \citet{PrataSparrow1983}, it is known that steady laminar flow in channels with periodic solid structures becomes periodically developed after some distance from the channel inlet.
At least, that is provided that the channel is long enough and fluid properties remain constant.
This means that the flow patterns, or streamlines, become similar around each solid structure, so that the velocity field is spatially periodic along the main flow direction.
When the array is sufficiently wide, also transversal flow periodicity will then occur at some distance from the channel's side walls.
In micro channels, the occurrence of periodically developed flow has recently been confirmed by means of micro-PIV measurements.
Experimental evidence for its occurrence in micro channels with arrays of circular and square cylinders appears specifically from the measurements conducted by \citet{RenferBrunschwiler2011} and \citet{XuPanWu2018}.

From a mathematical point of view, periodically developed flow is described by the periodic flow equations first formulated by \citet{Patankar1977}.
According to these equations, periodic flow is driven by a constant average pressure gradient, which can also be defined as the gradient of the macro-scale pressure  obtained by repeated volume-averaging of the original pressure field \citep{BuckinxBaelmans2015}.
As the periodic flow equations can be solved on a single unit cell of the array, they have been used extensively as an approximate flow model whenever an analysis of the full flow field through the entire array of solid structures is infeasible.
%Up to present, they have been an essential mathematical model
Therefore, they are still the most widely adopted modelling approach for numerically predicting the relationship between the mass flow rate and pressure drop in arrays of periodic solid structures, as well as ordered porous media.
For square and circular two-dimensional cylinders, we refer specifically to the works of \cite{EdwardsShapiro1990, Ghaddar1995, Koch1997, AmaralSoutoMoyne1997} and \cite{Martin1998}.
The latter have been extended by the works of \cite{ PapathanasiouDendy2001}  \cite{LasseuxAhmadi2011} and \cite{KhalifaPocherTilton2020}.

%In some studies, the solid structures representing %ordered porous media.

Given the relevance of the periodically developed flow regime for technological applications, as well as theoretical modelling purposes, a reliable prediction of the onset and extent of this regime in common channel geometries is of crucial importance.
In view of this, direct numerical simulation (DNS) of the full-scale flow field in the entire channel is expected to result in the most accurate prediction.
However, DNS requires a tremendous amount of computational resources, which often exceeds the capabilities of current high-performance supercomputing infrastructure.
In addition, it necessitates efficient highly-parallellized flow simulation software codes to keep the simulation time manageable.
For that reason, reduced-order models requiring less computational resources that allow for a reliable estimation based on limited a priori information deem necessary.

%based on limited a priori information deem most necessary,
%
% that allow for a reliable estimation based on limited a priori information deem most necessary, 
%
%
%Nevertheless , because 
%deem most necessary for  reliable estimation based on limited a priori information and

Despite this necessity, the onset and extent of the periodically developed flow regime have not been systematically investigated before.
As such, reference data is lacking even for simple geometries like arrays of square cylinders confined by rectangular channel walls.
As a matter of fact, very few full-scale simulations have been reported in the literature, so that the physical features of the developing flow upstream of the onset point of periodically developed flow are still poorly understood.
Therefore, the aim of the present work is to describe the features of the developing flow which occur immediately upstream of the region of periodically developed flow.
This flow region is further called the \textit{quasi-periodically developed flow region}. 
In particular, we aim to investigate the eigenvalues which characterize this regime in micro and mini channels containing arrays of equidistant square in-line cylinders. 
Moreover, we will explore the factors influencing the onset and end point of quasi-periodically developed flow in such channels by means of direct numerical simulation. 
Finally, we will give a  quantitative indication of the perturbation sizes that can be expected from direct numerical simulation. 
Hereto, a parabolic profile for the velocity at the channel inlet
is assumed, although it is rather serves as a reference for future engineering analyses, than that it represents a realistic inlet condition.

The present article is just a stub in its current form, and it will be completed in the near future.
Our main motivation for publishing the following results in their current state, is that they form the theoretical basis for the macro-scale descriptions of quasi-periodically developed flow recently elaborated by \citet{Buckinx2022} and \citet{Vangeffelen2023}.
Therefore, the current work is a direct step towards the assessment of the validity of the macro-scale descriptions of the periodically developed flow and heat transfer regimes presented in \citep{BuckinxBaelmans2015, BuckinxBaelmans2015b, BuckinxBaelmans2016}.
%In addition, we aim to provide numerical data focylinder arrays which are representative for common microfluidic devices.

Because the current article is not yet finalized, its proper context within the preceding literature on developing flow is still missing.
For that reason, we refer the reader to the literature reviews given by \citet{Sadri1997, Sadri2002, Sadri2002b}.
The latter works give a thorough analysis of quasi-developed Poiseuille flow, as well as solutions for the developing flow region in two-dimensional plate channels.
In that regard, our work can be seen as a generalization of those theoretical and numerical results towards channels with arrays of periodic solid structures.
 
The remainder of this article is structured as follows.
In section 2, we present a first mathematical description of quasi-periodically developed flow.
In section 3, we present our numerical results in their ad-hoc state, although without any discussion or conclusions.

\section{Description of Quasi-Periodically Developed Flow}
\label{sec: Description of Quasi-Periodically Developed Flow}
%\subsection{Definition and Description of Quasi-Periodically Developed Flow}
To construct a mathematical description of quasi-periodically developed flow, we introduce a perturbation approach similar to the one adopted by \cite{Sadri1997, Sadri2002}.
We consider a steady perturbation of the periodically developed velocity field $\boldsymbol{u}^{\star}$, and assume that this perturbation decays exponentially with the coordinate $x_1$ along the main flow direction $\boldsymbol{e}_1$.
The actual velocity field $\boldsymbol{u}$ upstream of the periodically developed flow region is then given by
\begin{equation}
\label{eq: quasi-periodically developed flow}
\boldsymbol{u} - \boldsymbol{u}^{\star} = \velamp \exp\left(-\lambda x_1 \right) \triangleq \velamp \exp\left(-\boldsymbol{\lambda} \boldsymbol{\cdot}  \boldsymbol{x}  \right) \,,
\end{equation} 
where $\boldsymbol{\lambda} \triangleq \lambda\boldsymbol{e}_1$ is the decay rate and $\velamp$ is the perturbation's amplitude, which varies with the position $\boldsymbol{x}$.
We remark that by its definition, the developed velocity field in (\ref{eq: quasi-periodically developed flow}) exhibits streamwise periodicity over the array of solid structures.
So, we have
\begin{equation}
\label{eq: flow periodicity main flow direction}
\boldsymbol{u}^{\star} (\boldsymbol{x}) = \boldsymbol{u}^{\star}(\boldsymbol{x} + n_1 \boldsymbol{l}_1)
\end{equation}
for some integer $n_1$, given that $\boldsymbol{l}_1$ denotes the lattice vector spanning a row of the array, parallel to the main flow direction.
Although the developed velocity field may also display transversal periodicity when the channel is wide enough, i.e.
\begin{equation}
\boldsymbol{u}^{\star} (\boldsymbol{x}) = \boldsymbol{u}^{\star}(\boldsymbol{x} + n_2 \boldsymbol{l}_2)
\end{equation}
for positions $\boldsymbol{x}$ sufficiently far away from the channel's side walls, this is in general not always the case.

Provided that the perturbation is small, i.e. $\Vert \velamp \Vert \ll \Vert \boldsymbol{u}^{\star} \Vert$, its amplitude $\velamp$ and decay rate $\boldsymbol{\lambda}$ must obey the following momentum and mass conservation equations:
\begin{subequations}
\label{eq: quasi-periodically developed flow equations}
\begin{align}
\rho_f \boldsymbol{u}^{\star}  \boldsymbol{\cdot} \boldsymbol{\nabla} \velamp +
\rho_f \velamp \boldsymbol{\cdot} \boldsymbol{\nabla} \boldsymbol{u}^{\star} 
&= 
- \boldsymbol{\nabla} \pressamp+ 
\mu_f \nabla^2  \velamp  +
\rho_f (\boldsymbol{u}_f^{\star}  \boldsymbol{\cdot} \boldsymbol{\lambda}) \velamp 
\\
& \qquad - 
2\mu_f \boldsymbol{\lambda}  \boldsymbol{\cdot} \boldsymbol{\nabla}  \velamp + 
\mu_f (\boldsymbol{\lambda}  \boldsymbol{\cdot} \boldsymbol{\lambda})  \velamp +
\boldsymbol{\lambda} \pressamp \, 
\label{eq: quasi-periodically developed flow equations: momentum}
\\
\boldsymbol{\nabla}\boldsymbol{\cdot}  \velamp  &= \boldsymbol{\lambda} \boldsymbol{\cdot} \velamp \,, 
\label{eq: quasi-periodically developed flow equations: continuity}
\\
\velamp(\boldsymbol{x}) &= \velamp(\boldsymbol{x} + n_1 \boldsymbol{l}_1) \,, \qquad
\pressamp(\boldsymbol{x}) = \pressamp(\boldsymbol{x} + n_1 \boldsymbol{l}_1)
\,, 
\label{eq: periodicity of velocity and pressure mode amplitudes}
\\
 \velamp (\boldsymbol{x}) &= 0 \qquad \mbox{for} ~~ \boldsymbol{x}  \in \Gamma_{0}  \,. 
\label{eq: quasi-periodically developed flow equations: no slip}
\end{align}
\end{subequations}
The former equations have been obtained from the Navier-Stokes equations by substituting the expression (\ref{eq: quasi-periodically developed flow}) for the velocity field, and by decomposing the pressure field as
\begin{equation}
\label{eq: quasi periodically developed pressure field}
p(\boldsymbol{x}) = \pressamp(\boldsymbol{x}) \exp\left(-\boldsymbol{\lambda} \boldsymbol{\cdot}  \boldsymbol{x}  \right) + \boldsymbol{\nabla }\mathrm{P}_{\text{dev}} \boldsymbol{\cdot}\boldsymbol{x} + p^{\star}(\boldsymbol{x}) \,.
\end{equation}
This decomposition is based on the notion that the pressure field in the periodically developed flow region consists of a periodic part $p^{\star}$ and linear contribution due to the constant average pressure gradient $\boldsymbol{\nabla }\mathrm{P}_{\text{dev}}$ \citep{Patankar1977}.
Therefore, the first term on the right-hand side of (\ref{eq: quasi periodically developed pressure field}) corresponds to the perturbation of the developed pressure field in response to the considered velocity perturbation (\ref{eq: quasi-periodically developed flow}). 
Equations (\ref{eq: quasi-periodically developed flow equations}) rely on a linearisation of the Navier-Stokes equations  with respect to the developed velocity field $\boldsymbol{u}^{\star}$.
The much smaller advection terms which are proportional to $\exp\left(-2 \lambda x_1 \right)$ have thus been neglected, consistent with our preposition that the asymptotic convergence of $\boldsymbol{u}$ towards $\boldsymbol{u}^{\star}$ is determined by a single exponential mode, $\exp\left(-\lambda x_1 \right)$.
This explains why equations (\ref{eq: quasi-periodically developed flow equations}) are linear both in $\velamp$ and $\pressamp$.

As indicated by (\ref{eq: periodicity of velocity and pressure mode amplitudes}), the velocity amplitude $\velamp$  inherits its streamwise periodicity from $\boldsymbol{u}^{\star}$ because the velocity perturbation is advected by the developed velocity field  according to the linearized momentum equation (\ref{eq: quasi-periodically developed flow equations: momentum}).
Therefore, also the pressure amplitude $\pressamp$ will be streamwise periodic.
Nevertheless, the amplitudes $\velamp$ and $\pressamp$ will not display transversal periodicity, even if the developed velocity field $\boldsymbol{u}^{\star}$ would do so.
The reason is that the velocity perturbation needs to satisfy just the no-slip condition at the channel walls $\Gamma_{0}$, as expressed by (\ref{eq: quasi-periodically developed flow equations: no slip}).
As such, there is no mechanism that will impose transversal periodicity on $\velamp$.

In order to obtain an equation for the decay rate $\boldsymbol{\lambda}$, we average the momentum equation (\ref{eq: quasi-periodically developed flow equations: momentum}) over $n_1$ rows of the array, to find that $\boldsymbol{\lambda}$ is the  solution of the quadratic equation
\begin{equation}
\label{eq: eigenvalue problem lambda}
\rho_f \boldsymbol{\lambda} \boldsymbol{\cdot} \left( \langle \velamp \boldsymbol{u}^{\star} \rangle +
\langle \boldsymbol{u}^{\star} \velamp \rangle
\right) + 
\langle \boldsymbol{n}_0 \boldsymbol{\cdot} \left(- \pressamp_f \boldsymbol{I} + \mu_f \boldsymbol{\nabla} \velamp_f \right) \delta_0 \rangle  + \mu_f \lambda^2 \langle \velamp \rangle + 
\boldsymbol{\lambda} \langle \pressamp \rangle
=0\,.
\end{equation}
Here, each term between the brackets of the volume-averaging operator $\langle \; \rangle$ is spatially constant.
Further, $\delta_0$ denotes the Dirac surface indicator \citep{BuckinxBaelmans2015, Gagnon1970} for the no-slip surface $\Gamma_0$, whose normal $\boldsymbol{n}_0$ is directed outward of the fluid and hence towards the solid.
The subscript $f$ reminds us of the fact that the integration of 
$\velamp$ and $\pressamp$ along $\Gamma_0$ has to be carried out only along the side in contact with the fluid.

Together with the flow equations (\ref{eq: quasi-periodically developed flow equations}), the quadratic equation (\ref{eq: eigenvalue problem lambda}) forms an eigenvalue problem which generalizes the Orr-Sommerfeld equation for quasi-developed Poiseuille flow \citep{Sadri1997, Sadri2002}.
Its solution consists of an infinite series of eigenvalues $\lambda$ and corresponding velocity and pressure modes, $\velamp$ and $\pressamp$, which specify the ways the developed flow is mathematically allowed to be perturbed according to the Navier-Stokes equations.
However, on physical grounds, we expect that only the mode corresponding to the smallest positive eigenvalue can occur upstream of the periodically developed flow region.
After all, the modes corresponding to larger eigenvalues are destined to vanish much faster over space, so that they are unlikely to form stable flow solutions that can be effectively observed.

Obviously, in order to solve the flow equations (\ref{eq: quasi-periodically developed flow equations}) and the associated eigenvalue problem (\ref{eq: eigenvalue problem lambda}), first the periodically developed flow field $\boldsymbol{u}^{\star}$ must be determined. 
This can be done by solving the steady periodically developed flow equations (\cite{Patankar1977}) on the same row(s) of the array:
\begin{subequations}
\label{eq: periodically developed flow equations}
\begin{align}
\rho_f \boldsymbol{\nabla} \boldsymbol{\cdot} \left(
\boldsymbol{u}^{\star} \boldsymbol{u}^{\star}  \right)  &= 
- \boldsymbol{\nabla}\mathrm{P}_{\text{dev}}
- \boldsymbol{\nabla} p^{\star} + 
\mu_f \nabla^2  \boldsymbol{u}^{\star}   \,, \\
\boldsymbol{\nabla}\boldsymbol{\cdot}  \boldsymbol{u}^{\star}  &= 0 \,, \\
\boldsymbol{u}^{\star} (\boldsymbol{x}) &= \boldsymbol{u}^{\star} (\boldsymbol{x} + n_1 \boldsymbol{l}_1) \,, \qquad
p^{\star}(\boldsymbol{x}) =p^{\star}(\boldsymbol{x} + n_1 \boldsymbol{l}_1)
\,, \\
\boldsymbol{u}^{\star} (\boldsymbol{x}) &= 0 \qquad \mbox{for} ~~ \boldsymbol{x}  \in \Gamma_{0}  \,. 
\label{eq: periodically developed flow equations: no-slip}
\end{align}
\end{subequations}
Here, we recognize again the periodicity condition (\ref{eq: flow periodicity main flow direction}), as well as the no-slip condition (\ref{eq: periodically developed flow equations: no-slip}
).
To solve the periodically developed flow equations (\ref{eq: periodically developed flow equations}), one can either impose $\boldsymbol{\nabla}\mathrm{P}_{\text{dev}}$ directly, or treat $\boldsymbol{\nabla}\mathrm{P}_{\text{dev}}$ as an unknown whose value should meet the actual bulk velocity $u_b$ through the channel 
\begin{align}
\label{eq: constraint volume averaged velocity over row}
\langle \boldsymbol{u}^{\star}\rangle &= u_b \boldsymbol{e}_1 \,.
\end{align}
This integral constraint is a consequence of the principle of mass conservation and the periodicity of the flow field.
If the relationship between $\boldsymbol{\nabla}\mathrm{P}_{\text{dev}}$ and $u_b$ or $\boldsymbol{U}$ is a one-to-one mapping, both of the latter approaches are equivalent.

It must be emphasized that the system of equations (\ref{eq: quasi-periodically developed flow equations}), (\ref{eq: eigenvalue problem lambda}) and (\ref{eq: periodically developed flow equations}) only has a unique solution when the additional constraints
\begin{align}
\label{eq: scaling constant quasi-periodically developed velocity mode}
\boldsymbol{e}_1 \boldsymbol{\cdot} \langle \velamp \rangle &= C_{\velamp} \,,
&& \langle p^{\star} \rangle = p_0
\end{align}
are imposed.
Yet, the scaling factor $C_{\velamp}$ and offset $p_0$ which determine the absolute values of the velocity field and pressure field depend on the way the flow develops and thus the specific inlet conditions for the channel flow.
As such, they are usually not known without information from a direct numerical simulation of the entire flow through the channel.
So, a reconstruction of the flow field from the bulk velocity $u_b$ or uniform macro-scale velocity $\boldsymbol{U}$ on $n_1$ rows of the array, via (\ref{eq: constraint volume averaged velocity over row}) and (\ref{eq: scaling constant quasi-periodically developed velocity mode}), is in practise achievable up to the unknown scaling factor $C_{\velamp}$.

It is interesting to note that the scaling factor $C_{\velamp}$ has no relation to the mass flow rate through the channel of height $H$ and width $W$, since the amplitude of the exponential velocity mode satisfies
\begin{equation}
\displaystyle \int_{0}^{H} \int_{0}^{W} \velampx(x_1, x_2, x_3) \, dx_2 \,dx_3 = 0
\end{equation}
for any section $x_1$.
This is a consequence of the fact that one can easily prove that only the periodic velocity part $\boldsymbol{u}^{\star}$ contributes to the mass flow rate \cite{BuckinxPhD2017}:
\begin{equation}
\displaystyle \frac{1}{WH} \int_{0}^{H} \int_{0}^{W} u_1^{\star}(x_1, x_2, x_3) \, dx_2 \,dx_3 = u_b \,.
\end{equation}

\section{Quasi-Periodically Flow in Arrays of In-Line Square Cylinders}
In order to show that preceding perturbations or modes, and hence  the region of quasi-periodically flow, effectively occur in channel flows, empirical evidence is required.
This empirical evidence is here obtained by means of DNS of full-scale channel flows.
%The data presented in the next figure has been obtained by means of direct numerical simulation of the flow, for the boundary conditions discussed in section \ref{sec: Channel Domain and Array Geometry}.
Each direct numerical simulation was performed on a regularly-sized mesh of about $140$ to $250$ million mesh cells, resulting in a typical computational time of about $20$ to $48$ hours on $13$ nodes of $36$ processors, for a total number of $2500$ to $7500$ discrete time steps of size $\Delta t = 0.08\ell_1/u_b$.
To verify the mesh-independence of the calculated flow field, each simulation was redone on a coarser mesh of about $180$ millon mesh cells, as well as two even coarser meshes of about $34$ and $88$ million mesh cells.

The next figures illustrate the onset point and extent of the quasi-periodically developed flow, as well as the corresponding exponential mode and eigenvalues for some channel flows through arrays of square in-line cylinders. 

\begin{figure}
\begin{center}
\includegraphics[scale=0.45]{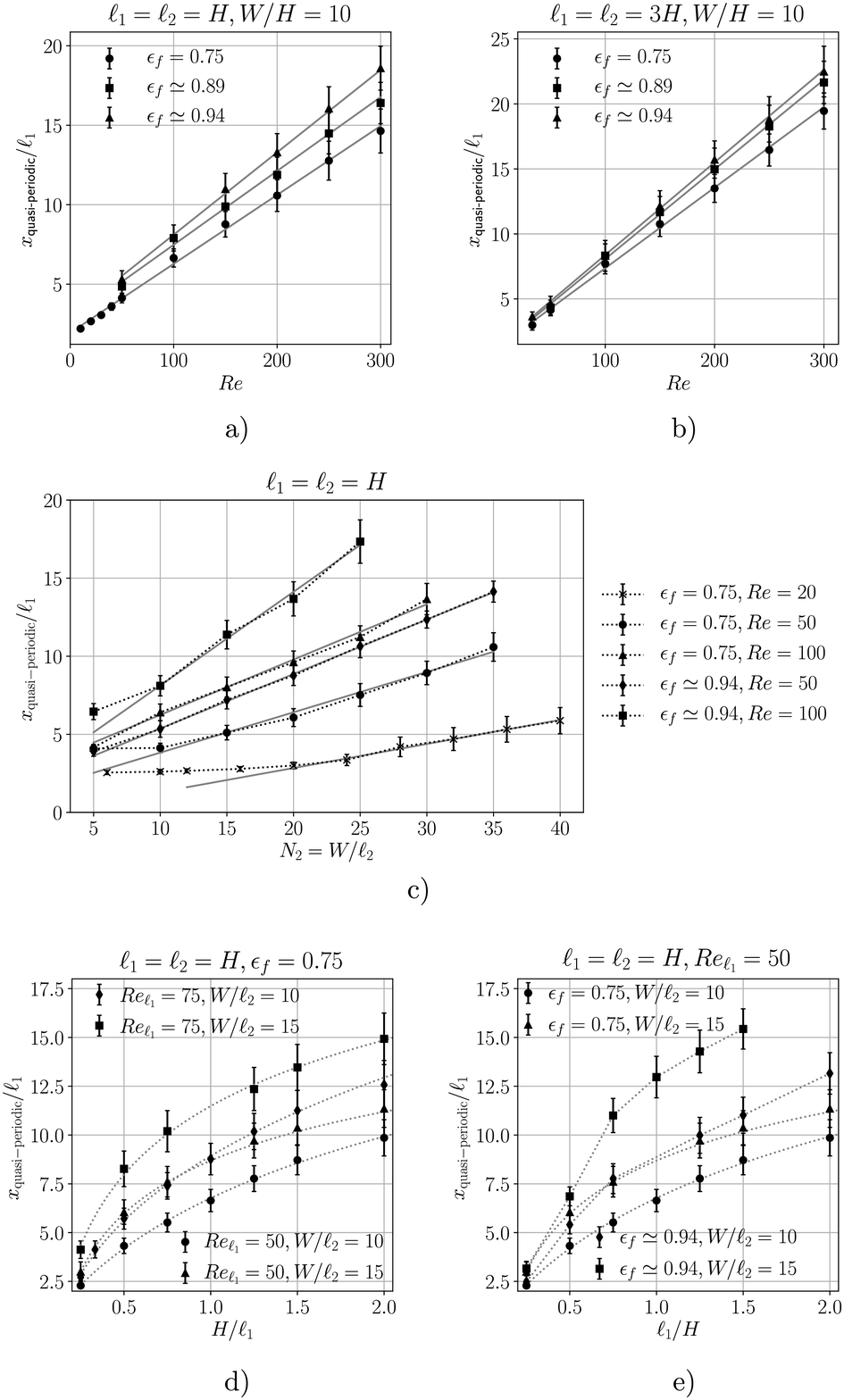}
\caption{Dependence of the onset point of quasi-periodically developed flow, $x_{\text{quasi-periodic}}$, on the Reynolds number $Re$ (a, b), the dimensionless channel width $W_2/\ell_2$ (c), and the dimensionless channel height $H/\ell_1$ (d, e), for different porosities $\epsilon_f$ of the cylinder array.
The Reynolds number $Re_{\ell_1}$ is based on the cylinder spacing $\ell_1=\ell_2$, whereas $Re$ is based on the double channel height $2H$.}
\label{fig: Onset point quasi-periodically flow}
\end{center}
\end{figure}

\begin{figure}
\begin{center}
\includegraphics[scale=0.45]{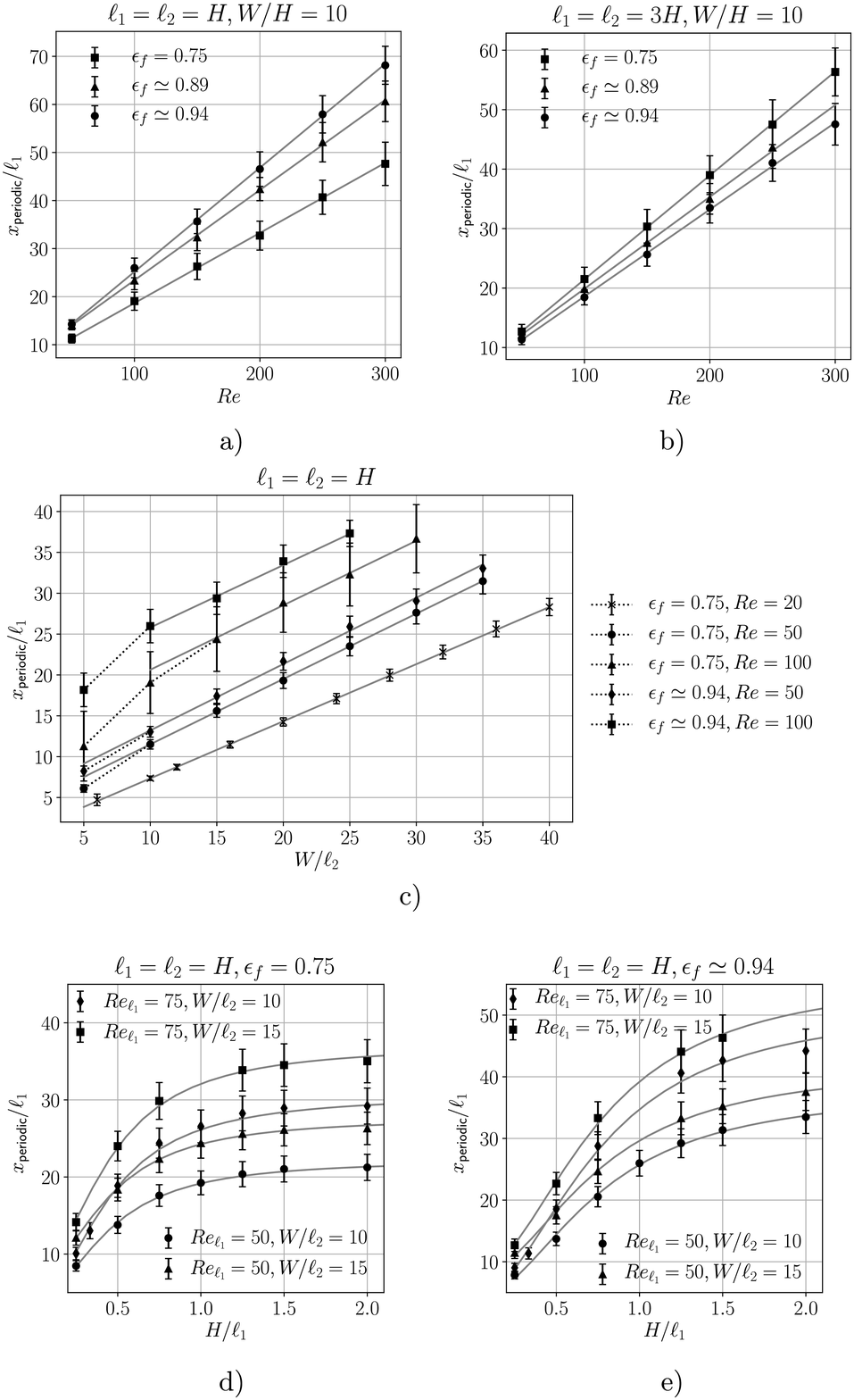}
\caption{Dependence of the onset point of periodically developed flow, $x_{\text{periodic}}$, on the Reynolds number $Re$ (a, b), the dimensionless channel width $W_2/\ell_2$ (c), and the dimensionless channel height $H/\ell_1$ (d, e), for different porosities $\epsilon_f$ of the cylinder array.
The Reynolds number $Re_{\ell_1}$ is based on the cylinder spacing $\ell_1=\ell_2$, whereas $Re$ is based on the double channel height $2H$.}
\label{fig: Onset point periodically developed flow}
\end{center}
\end{figure}

\begin{figure}
\begin{center}
\includegraphics[scale=0.45]{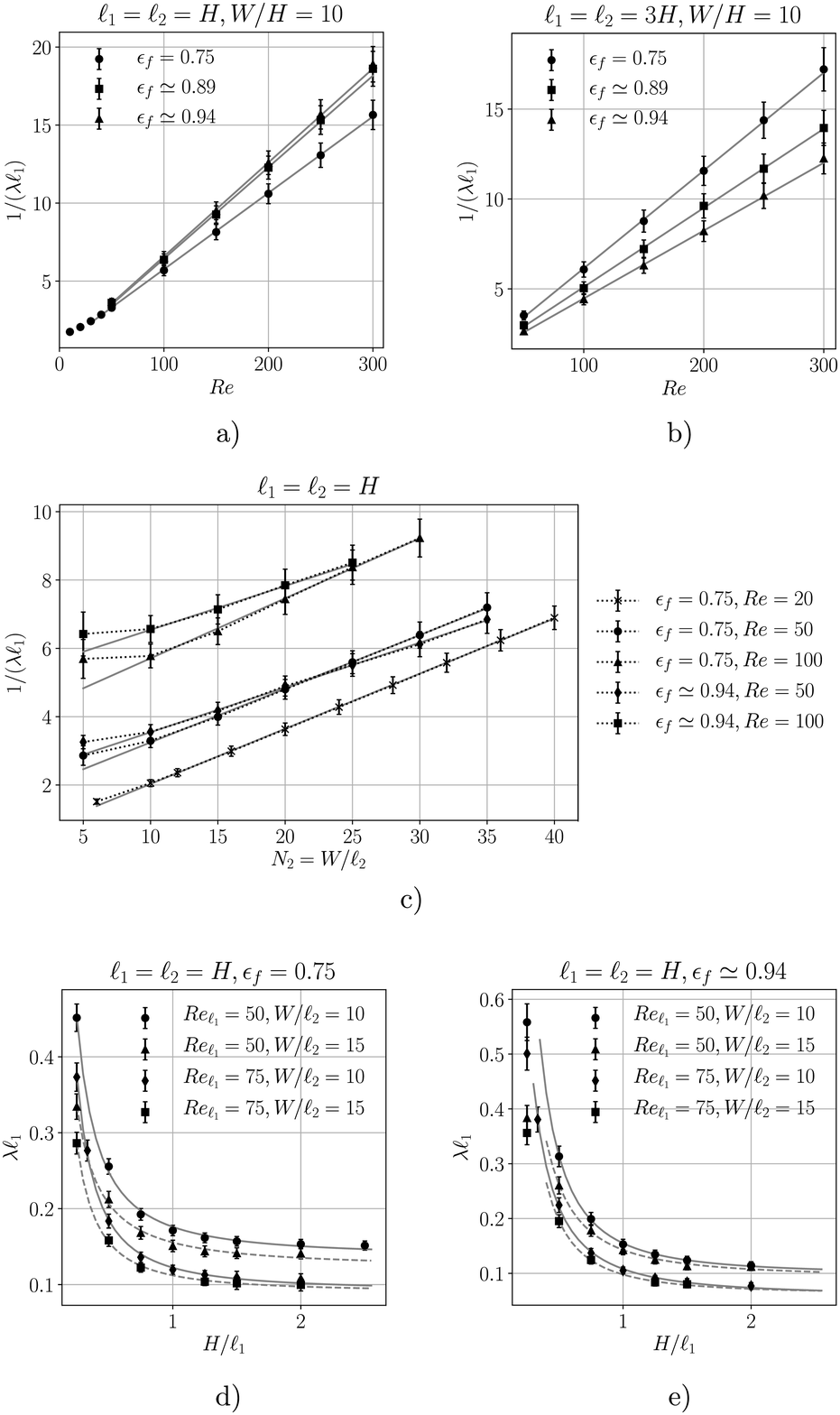}
\caption{Dependence of the eiegnvalues of periodically developed flow, $x_{\text{periodic}}$, on the Reynolds number $Re$ (a, b), the dimensionless channel width $W_2/\ell_2$ (c), and the dimensionless channel height $H/\ell_1$ (d, e), for different porosities $\epsilon_f$ of the cylinder array.
The Reynolds number $Re_{\ell_1}$ is based on the cylinder spacing $\ell_1=\ell_2$, whereas $Re$ is based on the double channel height $2H$.}
\label{fig: Eigenvalues of quasi-periodically developed flow}
\end{center}
\end{figure}

\begin{figure}
\begin{center}
\includegraphics[scale=0.45]{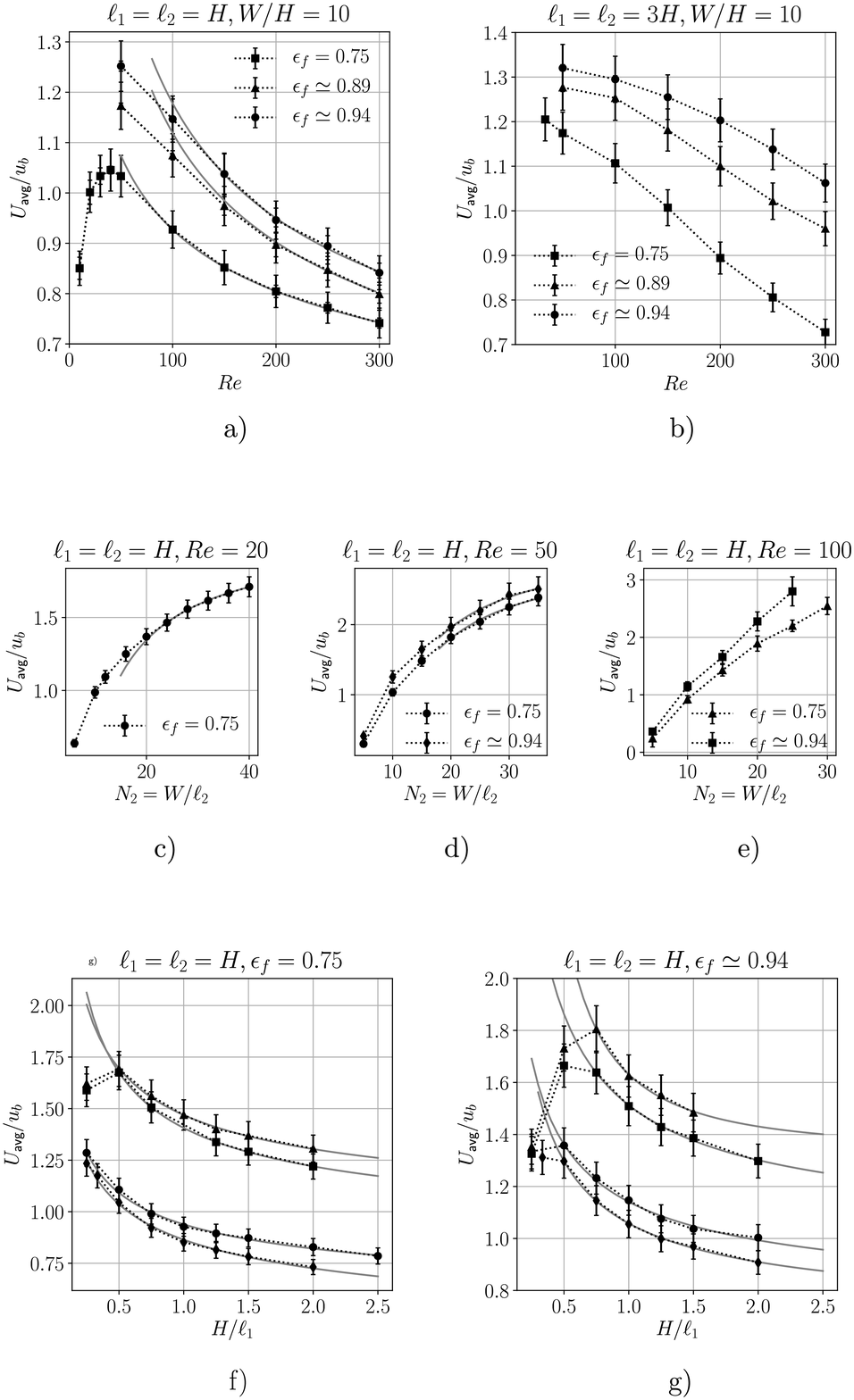}
\caption{Dependence of the perturbation sizes of periodically developed flow, $x_{\text{periodic}}$, on the Reynolds number $Re$ (a, b), the dimensionless channel width $W_2/\ell_2$ (c, d, e), and the dimensionless channel height $H/\ell_1$ (f, g), for different porosities $\epsilon_f$ of the cylinder array.
The Reynolds number $Re_{\ell_1}$ is based on the cylinder spacing $\ell_1=\ell_2$, whereas $Re$ is based on the double channel height $2H$.}
\label{fig: Amplitudes of quasi-periodically developed flow}
\end{center}
\end{figure}

\section*{Author Contribution}
G.B. contributed to the entire conceptualization, investigation,  methodology and writing of the present work, including all theoretical derivations, numerical results, data analysis, and software development. 
A.V. contributed by independently verifying the numerical data, and provided additional data for Reynolds numbers below 25, as well as the data for aspect ratios above 25.

\section{\label{sec:acknowledgements}Acknowledgements}
The work presented in this paper was partly funded by the Research Foundation — Flanders (FWO) through the post-doctoral project grant 12Y2919N of G. Buckinx.
%, and partly by the Flemish Institute for Technological Research (VITO) through the Ph.D. grant 1810603 of A. Vangeffelen. 
The VSC (Flemish Supercomputer Center), funded by the Research Foundation - Flanders (FWO) and the Flemish Government, provided the resources and services used in this work. 

\newpage
\clearpage

%%\nocite{*}
\bibliography{References}% Produces the bibliography via BibTeX.

\end{document}